\documentclass[journal]{IEEEtran}
\usepackage{amsmath}
\usepackage{amsthm} 
\usepackage{stfloats,float}
\usepackage{bm}
\usepackage{algorithm,algorithmic}
\usepackage{epsfig,subfigure,cite,amssymb,flushend,amsfonts,color,xcolor,multirow}%
\usepackage{cuted}
\usepackage{booktabs}
\usepackage{tabularx}
\usepackage{hyperref} 
\usepackage{graphicx} 
\begin{document}
%
\title{Utilizing Improper Gaussian Signaling for Downlink Rate-Splitting Multiple Access Systems with Imperfect Successive Interference Cancellation}

%
\author{Wanting Shi, Hao Cheng, Zhe Li, Yili Xia, \emph{Member}, \emph{IEEE}, and Wenjiang Pei
\thanks{W. Shi, Y. Xia, and W. Pei are with the School of Information Science and Engineering, Southeast University, Nanjing 210096, China. (e-mail:\{wanting\_shi,yili\_xia,wjpei\}@seu.edu.cn).}
\thanks{H. Cheng is with the School of Internet of Things, Nanjing University of Posts and Telecommunications, Nanjing 210003, China. (e-mail:haocheng@njupt.edu.cn).}
\thanks{Z. Li is with the School of Electronic and Information Engineering, Soochow University, Suzhou 215006, China. (e-mail:lizhe@suda.edu.cn).}
}
\maketitle
\begin{abstract}
To mitigate the residual interference from imperfect successive interference cancellation (SIC) in Rate-Splitting Multiple Access (RSMA), this paper incorporates improper Gaussian signaling (IGS) into the downlink RSMA framework. 
Unlike existing RSMA--IGS works that embed signal impropriety within I/Q-imbalanced frameworks, we show that IGS alone effectively counters SIC-induced residual interference. 
A basic Single-Input Single-Output (SISO) setup is considered with IGS on the common stream and proper Gaussian signaling (PGS) on the private streams, and it is analyzed in three parts. First, we show that the private rate maximization is achieved at the maximum signal impropriety. Second, we derive closed-form optimal solutions with rigorous monotonicity conditions for the common rate maximization. Third, to address the non-convex sum rate maximization, we develop a soft actor-critic (SAC) based algorithm that efficiently explores the solution space. Theoretical analysis and numerical results demonstrate that IGS consistently surpasses PGS, and its performance advantage becomes more pronounced when the SIC imperfection becomes severe.
\end{abstract}
\IEEEpeerreviewmaketitle
\vspace{-0.2cm}
\section{Introduction}
Rate-Splitting Multiple Access (RSMA) has emerged as a pivotal technology for next-generation wireless networks, renowned for its robustness and ability to manage multi-user interference~\cite{mao2018rate, Mao_Yijie_2022}. The core principle of RSMA involves splitting a user's message into a common part and a private part. The common parts are superimposed into a single stream decoded by all users, while the private parts are decoded individually, treating residual interference as noise. This structure allows for RSMA a seamlessly transition between fully orthogonal (e.g., SDMA) and fully non-orthogonal (e.g., NOMA) transmission schemes, thereby achieving superior spectral efficiency and enhanced user fairness.

However, a significant body of RSMA research relies on the ideal assumption of perfect successive interference cancellation at the receivers~\cite{Soleymani_2023_TVT}. In practice, SIC is inherently imperfect due to hardware limitations, channel estimation errors, and finite blocklength effects~\cite{Cheng_Hao_2022}. 
Existing mitigation strategies, such as power control~\cite{Xu_Yongjun2024} and RIS deployment~\cite{WANG2025102761}, primarily focus on the system-level resource allocation and the hardware-assisted optimization. However, these approaches fail to address a fundamental limitation inherent in the signal statistics: the conventional assumption of proper Gaussian signaling (PGS). By restricting the pseudo-covariance to zero, PGS fundamentally forfeits the additional degrees of freedom residing in the complementary covariance domain.

This observation motivates the consideration of improper Gaussian signaling (IGS), which relaxes the pseudo-variance constraint of PGS to exploit the full covariance structure of complex signals. By shaping the pseudo-variance, IGS can selectively attenuate the interference footprint at undesired receivers, offering inherent robustness against residual interference without additional hardware or spectrum resources. While IGS has shown promise in interference channels~\cite{ZengYong_2013, Cheng_Hao_2022} and cognitive radio~\cite{Soleymani_2023_TVT}, its potential in RSMA systems—particularly under imperfect SIC—remains entirely unexplored. A key challenge lies in the coupled interplay between the improper degree of the common stream and the private-stream power allocation, which jointly determine both the decodability of the common message and the residual interference seen after SIC.

To address this gap, we investigate the integration of IGS into a downlink RSMA system impaired by imperfect SIC. 
While prior literature utilizes impropriety to offset hardware distortions~\cite{Soleyman_TGCN_2022}, our analysis reveals that IGS itself can effectively combat residual interference arising from imperfect SIC. 
More importantly, in contrast to recent hybrid RSMA--IGS studies under imperfect SIC that resort to Multiple-Input Multiple-Output (MIMO) architectures and opaque numerical optimization~\cite{Xu_TCOM_2026}, our single-input single-output (SISO) framework deliberately trades architectural complexity for analytical tractability, yielding closed-form expressions that transparently reveal how the degree of impropriety $\kappa$ shapes each rate component. Specifically, we prove that $\kappa^{\star} = 1$ for private rate maximization, derive explicit analytical expressions for the optimal $\kappa$ in common rate maximization with rigorous monotonicity conditions, and develop a soft actor-critic (SAC) algorithm for the remaining non-convex sum rate problem. Numerical results confirm that IGS consistently outperforms PGS, and the gain widens as the SIC imperfection increases. 
\vspace{-0.2cm}
\section{System model}
\subsection{Preliminaries for Improper Gaussian Signaling}
A complex Gaussian random variable $s$ is characterized by its variance $\sigma_s^2 = \mathbb{E}[|s|^2]$ and pseudo-variance $\widetilde{\sigma}_s^2 = \mathbb{E}[s^2]$~\cite{Schreier_Book}. An improper Gaussian signal relaxes the constraint $\widetilde{\sigma}_s^2 = 0$ of PGS, introducing an additional design degree of freedom quantified by the degree of impropriety $\kappa=|\widetilde{\sigma}_s^2|/\sigma_s^2\in[0,1]$. This extra dimension has been shown to enhance interference management~\cite{ZengYong_2013,Cheng_Hao_2022,Soleymani_2023_TVT,Jin_Honglei_2025}.
\subsection{RSMA with IGS}
Quantifying the impact of IGS on RSMA system performance poses unique challenges due to the coupled constraints. To tackle this issue, we establish a fundamental two-user SISO RSMA framework to derive analytical expressions for performance characterization.
In this model, the common stream symbol $W_c$ is transmitted using IGS, whereas the private stream symbols $W_{1}$ and $W_{2}$ employ conventional PGS. 
For simplicity and without loss of generality, we order the users such that the channel gains satisfy $|h_1|\geqslant |h_2|$~\cite{Jin_Honglei_2025}, which streamlines the analysis while preserving generality for arbitrary channel realizations.

The received signal at user-$k$ is given by
\begin{align}
    y_{k} &= h_{k}\sqrt{p_{c}}x_{c} + \underbrace{h_{k}\sqrt{p_{k}}x_{k} + h_{k} \sqrt{p_{j}}x_{j} + n_{k}}_{z_{k}^{c}}, 
\end{align}
where $j,k\in\{1,2\}$, and $j\neq k$. $x_{c}$ denotes the common signal, using IGS with variance $\sigma_{x_c}^2$ and pseudo-variance $\tilde{\sigma}_{x_c}^2$, and its improper coefficient $\kappa = |\tilde{\sigma}_{x_c}^2|/\sigma_{x_c}^2$. The private signal $x_{k}$ is modeled as PGS with unit variance. The term $n_{k}$ is the additive white Gaussian noise with zero mean and variance $\sigma^2$, and the aggregated interference-plus-noise (IPN) term for the common stream is denoted by $z_{k}^{c}$.

Assuming imperfect SIC, the private signal for user-$k$ can be expressed as
\begin{align}
    y_{k}^p &= h_{k}\sqrt{p_{k}}x_{k} + \underbrace{\lambda h_{k}\sqrt{p_{c}}x_{c} + h_{k} \sqrt{p_{j}}x_{j} + n_{k}}_{z_{k}}
\end{align}
where $\lambda \in [0,1]$ models the degree of imperfect SIC, with a larger $\lambda$ indicating more residual interference. $z_{k}$ denotes the residual IPN term for the private stream. Based on this, the private rate for user-$k$ is derived as
\begin{align}
    &R_{k} = \frac{1}{2} \log_{2} \left( \frac{\sigma^4_{y_{k}^p} - |\widetilde{\sigma}_{y_{k}^p}^2|^2}{\sigma_{z_{k}}^4 - |\widetilde{\sigma}_{z_{k}}^2|^{2}} \right) \nonumber \\
    &= \frac{1}{2}\log_{2} \Biggl(1 + \frac{p_k\Gamma_k(2\lambda^2 p_c \Gamma_k + p_k\Gamma_k + 2p_{j}\Gamma_k + 2 )}{(\lambda^2 p_c \Gamma_k  + p_{j} \Gamma_k + 1)^2 - \lambda^4 \kappa^2 p_c^2 \Gamma_k^2} \Biggr),
    \label{eq:R_k}
\end{align}
where $ \Gamma_k = |h_k|^2/ \sigma ^2$, and $k\in\{1,2\}$ denotes the channel-to-noise ratio (CNR) for user-$k$.

Similarly, the common rate for user-$k$ is given by
\begin{align}
    R_{c,k}  &= \frac{1}{2} \log_{2} \left( \frac{\sigma^4_{y_k} - |\widetilde{\sigma}_{y_k}^2|^2}{\sigma_{z_{k}^{c}}^4 - |\widetilde{\sigma}_{z_{k}^{c}}^2|^{2}} \right)  \nonumber\\
    &= \frac{1}{2} \log_{2}  \biggl( \frac{[(p_c + p_1 + p_2)\Gamma_k+1]^2-\kappa^2 p_c^2 \Gamma_k^2}{[(p_1 + p_2)\Gamma_k+1]^2} \biggr).
    \label{eq:R_ck}  
\end{align}
It can be observed that $R_k$ increases monotonically with $\kappa$, while $R_{c,k}$ decreases monotonically with $\kappa$. This is because the denominator of $R_k$ in \eqref{eq:R_k}, $(\lambda^2 p_c \Gamma_k + p_{j}\Gamma_k + 1)^2 - \lambda^4 \kappa^2 p_c^2 \Gamma_k^2$, strictly decreases with $\kappa$, while the numerator remains independent of $\kappa$. Conversely, $R_{c,k}$ in \eqref{eq:R_ck} has a numerator that strictly decreases with $\kappa$ while the denominator is $\kappa$-independent.

To ensure that the common stream $x_c$ is successfully decoded by all users, the total stream rate $R_c$ must satisfy the worst-case user constraint, that is, $R_{c} = \min \{R_{c,k}\}$.
\section{Achievable Rate Analysis}
\subsection{Achievable Rate Maximization of the Private Streams}
\label{Private Streams}
In RSMA systems, private messages are designed to be decoded solely by their targeted users, offering a natural advantage for physical layer security. We first investigate how the degree of signal impropriety $\kappa$ enhances the maximum achievable private rate.

This optimization problem on the sum of private rates $R_1$ and $R_2$ is formulated as follows
\begin{align}
    \max_{\kappa,p_c,p_1,p_2} \ &R_{1}+R_{2},  \label{eq:Max_Private_Streams} \\
    s.t. \          &p_c \geqslant \tau_{SIC},   
                    \tag{\ref{eq:Max_Private_Streams}{a}} \label{eq:Max_Private_Streams_a} \\
                    &p_c + p_1 + p_2 \leqslant P,  
                    \tag{\ref{eq:Max_Private_Streams}{b}} \label{eq:Max_Private_Streams_b} \\
                    &0 \leqslant \kappa \leqslant 1,    
                    \tag{\ref{eq:Max_Private_Streams}{c}} \label{eq:Max_Private_Streams_c} 
\end{align}
where $P$ represents the total transmit power budget at the BS, and $\tau_{SIC}$ is the lower bound on the common stream power $p_c$ required to guarantee feasible decoding of the common message for the subsequent SIC procedure. Notably, this rate maximization enhances the security of RSMA systems and is frequently studied in various non-convex resource optimization problems~\cite{Jin_Honglei_2025,Salem_2023_TWC}.

\emph{Theorem 1}: \label{Theorem_1} With imperfect SIC, the objective in \eqref{eq:Max_Private_Streams} is maximized when the common stream exhibits maximum impropriety, i.e., $\kappa^{\star}=1$.

\emph{Proof}: As discussed in Section II, both $R_1$ and $R_2$ strictly increase with $\kappa$, and so does $R_1 + R_2$.
Moreover, the constraints in \eqref{eq:Max_Private_Streams_a}--\eqref{eq:Max_Private_Streams_c} involve only $p_c$, $p_1$, $p_2$, and $0 \leqslant \kappa \leqslant 1$, without any coupling between $\kappa$ and the power variables. Hence, the feasible region of $(p_c, p_1, p_2)$ is independent of $\kappa$, and increasing $\kappa$ always boosts the sum of private rates, $R_1+R_2$. The optimum is therefore attained at $\kappa^{\star} = 1$. $\hfill\blacksquare$

Taking the derivative of $R_k$ with respect to $p_c$ yields
\begin{align}
    \frac{d(R_{k})}{d(p_c)} &= \frac{\lambda^2 p_{k} \Gamma_{k}^2\left(A_2 - A_1A_3 \right)}{u A_2^2\ln 2},
\end{align}
where
\begin{align}
   A_1 &= 2\lambda^2 p_c \Gamma_{k} + p_{k}\Gamma_{k} + 2p_{j}\Gamma_{k} + 2,  \\
   A_2 &= (\lambda^2 p_c \Gamma_{k}  + p_{j} \Gamma_{k} + 1)^2 - \lambda^4 \kappa^2 p_c^2 \Gamma_{k}^2,    \\
   A_3 &= \lambda^2 p_c \Gamma_{k}  + p_{j} \Gamma_{k} + 1 -\lambda^2 \kappa^2 p_c\Gamma_{k},  \\
   u &= 1 + \frac{p_{k}\Gamma_{k} A_1}{A_2}.
\end{align}
It follows that $R_{k} = 1/2 \log_{2}u$.

Since $u$, $A_1$, $A_2$, $p_k$, and $\Gamma_k$ are all positive, the monotonicity of $R_k$ with respect to $p_c$ is determined by the term $A_2 - A_1 A_3$. When $\kappa=1$, this term is negative, establishing that $R_1 + R_2$ decreases with $p_c$. Thus, the optimal $p_c^\star$ achieves equality in \eqref{eq:Max_Private_Streams_a}: $p_c^{\star} = \tau_{SIC} $. 
%
%
The Lagrangian function for the optimization problem is constructed by introducing the non-negative Lagrange multipliers $\mu$, $\nu_1$ and $\nu_2$, which is given by
\begin{align}
    L(p_1,p_2,\mu,\nu_1,\nu_2) &= R_1 + R_2 + \mu(p_1+p_2+ \tau_{SIC}-P) 
     \nonumber \\
    + \nu_1 p_1 + \nu_2 p_2.  
\end{align}
An investigation on the first-order Karush-Kuhn-Tucker (KKT) conditions demonstrates that at the optimal power allocation, it must hold that $\partial (R_{1}+R_{2})/\partial p_1 = \partial (R_{1}+R_{2})/\partial p_2$. This result is analogous to that obtained by employing the water-filling principle, where the marginal gain in the sum rate with respect to power is equalized for both the users~\cite{Yang_Zhaohui_2021}.
\subsection{Achievable Rate Maximization of the Common Streams}
\label{Common Streams}
The common message underpins system fairness and SIC decodability, necessitating a sufficiently high common rate $R_c$. We therefore maximize the common rate while enforcing a minimum private rate $R_{\min}$.

The optimization problem is formulated as
\begin{align}
    \max_{\kappa,p_c,p_1,p_2} \ R_{c},   \label{eq:Max_Common_Streams}  \\
    s.t. \          &R_1, R_2 \geqslant R_{\min},      
                    \tag{\ref{eq:Max_Common_Streams}{a}} \label{eq:Max_Common_Streams_a}\\
                    &p_c \geqslant \tau_{SIC}, 
                    \tag{\ref{eq:Max_Common_Streams}{b}} \label{eq:Max_Common_Streams_b} \\
                    &p_c + p_1 + p_2 \leqslant P,  
                    \tag{\ref{eq:Max_Common_Streams}{c}} \label{eq:Max_Common_Streams_c} \\
                    &0 \leqslant \kappa \leqslant 1,    
                    \tag{\ref{eq:Max_Common_Streams}{d}} \label{eq:Max_Common_Streams_d} 
\end{align}
where $R_{\min}\triangleq \mathbb{E} [R_s]+\mathbb{E} [\max\{R_{i\to k}^p\}]$ denotes the simplified secrecy rate threshold~\cite{Salem_2023_TWC}.

We observe that $R_{c,1}$ and $R_{c,2}$ share identical functional forms and depend solely on the total power allocated to private streams ($p_1 + p_2$), rather than the specific power distribution between $p_1$ and $p_2$. Under the channel condition $|h_1| > |h_2|$, which implies that $\Gamma_1 > \Gamma_2$, it follows that $R_{c,1} > R_{c,2}$. Consequently, the original objective $R_c = \min\{R_{c,1}, R_{c,2}\}$ in \eqref{eq:Max_Common_Streams} simplifies to maximizing $R_{c,2}$.

To handle non-convexity, we first fix $p_1,p_2$ and study the degree of signal impropriety $\kappa$. At least one private rate constraint must be binding at the optimum, otherwise reallocating power from private to common streams would further increase $R_{c,2}$. Thus, we consider either $R_1=R_{\min}$ or $R_2=R_{\min}$.

Through a few algebraic manipulations, the binding rate constraint $R_{k} = R_{\min}$ in~\eqref{eq:Max_Common_Streams_a}, where $k\in\{1,2\}$, yields
\begin{align}
    B_1 p_c^2 + B_2 p_c + B_3 = 0,
    \label{eq:quadratic_pc}
\end{align}
where $S = 2^{2R_{\min}} - 1$ and
\begin{align}
    B_1 &= S \lambda^4 \Gamma_k^2 (1 - \kappa^2),
    \nonumber  \\
    B_2 &= 2\lambda^2 \Gamma_k \left[S (p_{j} \Gamma_k + 1) -  p_{j}\Gamma_k\right],
    \nonumber  \\
    B_3 &= S (p_{j} \Gamma_k + 1)^2 - p_k\Gamma_k(p_k\Gamma_k + 2p_{j}\Gamma_k + 2).
    \nonumber
\end{align}
Since $B_2$ and $B_3$ are $\kappa$-free while $B_1>0$ decreases with $\kappa$, the unique positive root is given by
\begin{align}
    p_c(\kappa) = \frac{-B_2 + \sqrt{B_2^2 - 4 B_1 B_3}}{2 B_1},
    \label{eq:solution_pc}
\end{align}
which is monotonically increasing in $\kappa$. A higher $\kappa$ reduces private-stream interference, and enables a higher power allocation for $p_c$ to compensate the degradation of common rate, $R_{c,2}$.

Equation \eqref{eq:quadratic_pc} can be alternatively expressed as:
\begin{align}
    \kappa^2 &=  \frac{ (\lambda^2 p_c \Gamma_k + p_{j} \Gamma_k +1)^2}{ \lambda^4 p_c^2 \Gamma_k^2}
    \nonumber\\
    &- \frac{p_k(2\lambda^2 p_c \Gamma_k + p_k\Gamma_k + 2p_{j}\Gamma_k + 2 ) }{ S \lambda^4 p_c^2 \Gamma_k }.
    \label{eq:solution_kappa}
\end{align}
In Case 1 with $R_1 = R_{\min}$ binding, substituting \eqref{eq:solution_kappa} yields
\begin{small}
    \begin{align}
        R_{c,2} &= \tfrac{1}{2}\log_{2}\biggl(1+\tfrac{p_1\Gamma_2^2(p_1\Gamma_1+2p_2\Gamma_1+2)}{S\lambda^4\Gamma_1}
        -\tfrac{(p_2\Gamma_1+1)^2\Gamma_2^2}{\lambda^4\Gamma_1^2}  \nonumber \\
        &+\tfrac{2[S\lambda^2\Gamma_1^2 C_1+p_1\Gamma_1\Gamma_2-S(p_2\Gamma_1+1)\Gamma_2]p_c(\kappa)\Gamma_2}{S\lambda^4\Gamma_1^2 C_1^2}\biggr),
        \label{eq:R_c2_Substitute_kappa1}
    \end{align}
\end{small}
where $C_1 = (p_1+p_2)\Gamma_1 + 1$. The monotonicity of $R_{c,2}$ with respect to $\kappa$ is governed by
\begin{align}
    \mathcal{M}_1 = S\lambda^2\Gamma_1^2 C_1 + p_1\Gamma_1\Gamma_2 - S(p_2\Gamma_1+1)\Gamma_2.
    \label{positivity_dRc2_dkappa1}
\end{align}
Similarly, in Case 2 with $R_2 = R_{\min}$ active, we have
\begin{small}
    \begin{align}
        R_{c,2}
        &= \tfrac{1}{2} \log_{2} \biggl( 1 + \tfrac{p_2\Gamma_2(p_2\Gamma_2+2p_1\Gamma_2)-S(p_1\Gamma_2+1)^2}{S \lambda^4 C_2^2}  \nonumber \\
        +& \tfrac{\big[S\lambda^2 C_2 + 2p_2\Gamma_2 - 2S(p_1\Gamma_2 + 1) \big]p_c(\kappa)\Gamma_2}{S\lambda^4C_2^2} \biggr)
        \label{eq:R_c2_Substitute_kappa}
    \end{align}
\end{small}
where $C_2 = (p_1+p_2)\Gamma_2 + 1$. The monotonicity of $R_{c,2}$ with respect to $\kappa$ is determined by
\begin{align}
    \mathcal{M}_2 = S\lambda^2 C_2 + 2p_2\Gamma_2 - 2S(p_1\Gamma_2 + 1).
    \label{positivity_dRc2_dkappa2}
\end{align}
Recall that $p_c(\kappa)$ is given in \eqref{eq:solution_pc} and that the total power constraint imposes $p_c(\kappa)\leqslant D\triangleq P-p_1-p_2$. 
Since $p_c(\kappa)$ is monotonically increasing in $\kappa$, this reduces to the conditions $p_c(0) \leq D$ and $p_c(1) \geq \tau_{\text{SIC}}$.

When feasible, the optimal $\kappa$ follows the same piecewise form in both the cases, determined by the sign of the corresponding monotonicity condition. Let $\mathcal{M}$ denote this condition: $\mathcal{M}=\mathcal{M}_1$ in Case 1 ($R_1=R_{\min}$) and $\mathcal{M}=\mathcal{M}_2$ in Case 2 ($R_2=R_{\min}$). Then,
\begin{align}
    \kappa^{\star} =
    \begin{cases}
    1, & \mathcal{M} > 0  \ \text{and}  \ p_c(1) \leq D,   \\
    \sqrt{1+\dfrac{B_2 D + B_3}{S\lambda^4 \Gamma_2^2 D^2}}, & \mathcal{M} > 0 \ \text{and} \ p_c(1) > D,   \\
    0, & \mathcal{M} \leq 0,
    \end{cases}
    \label{kappa_optimal}
\end{align}

\subsection{Achievable Rate Maximization of Sum Rate}
\label{sum rate}
The sum rate maximization of a two-user RSMA system can be formulated as
\begin{align}
    \max_{\kappa,p_c,p_1,p_2} \ R_{tot},   \label{eq:Max_sumrate}  \\
    s.t. \          &R_{1}, R_{2} \geqslant R_{\min},      
                    \tag{\ref{eq:Max_sumrate}{a}} \label{eq:Max_sumrate_a}\\
                    &p_c \geqslant \tau_{SIC},  
                    \tag{\ref{eq:Max_sumrate}{b}} \label{eq:Max_sumrate_b} \\
                    &p_c + p_1 + p_2 \leqslant P, 
                    \tag{\ref{eq:Max_sumrate}{c}} \label{eq:Max_sumrate_c} \\
                    &0 \leqslant \kappa \leqslant 1,    
                    \tag{\ref{eq:Max_sumrate}{d}} \label{eq:Max_sumrate_d} 
\end{align}
where $R_{tot}$ denotes the sum rate of users.

Problem \eqref{eq:Max_sumrate} is highly non-convex due to the coupling among system variables, which motivates the use of the SAC algorithm. 
As a stable and sample-efficient off-policy algorithm, SAC learns its policy by balancing reward maximization with an entropy bonus, effectively preventing premature convergence to poor local optima~\cite{haarnoja2018SAC}.
The problem is modeled as a Markov decision process with state $\mathcal{S} = \{h_1, h_2, \lambda, \tau_{\text{SIC}}, \sigma^2, P\}$, action $\mathcal{A} = \{p_c, p_1, p_2, \kappa\}$, and reward $r(\mathcal{S},\mathcal{A}) = R_{\text{tot}} - \psi U_t$, where $U_t =\sum_{k = 1}^{K} \max[R_{\min} - R_{k}, 0]$ quantifies the total minimum-rate violation and $\psi > 0$ is a penalty weight that discourages the agent from selecting power allocations that violate the rate constraints.
The cumulative discounted reward is $R_t = \sum_{t'=t}^T \gamma^{t'} r_{t'}$ with a discount factor $\gamma \in [0,1)$. The SAC training procedure, which jointly updates the actor, critics, and temperature parameter $\alpha$, is summarized in Algorithm~\ref{SAC_algorithm}.
\begin{algorithm}[t]
    \caption{SAC Algorithm for Sum Rate Maximization}
    \label{SAC_algorithm}
    \begin{algorithmic}[1]
        \STATE Initialize actor network $\pi_{\phi}$, critic networks $Q_{\theta_1}$ and $Q_{\theta_2}$, target networks $Q_{\bar{\theta}_1}$ and $Q_{\bar{\theta}_2}$, temperature $\alpha$, target update rate $\tau$, and replay buffer $\mathcal{D}$.
        \FOR{episode $= 1$ to $M$}
            \STATE Get initial state $\mathbf{s}$
            \FOR{each step} 
                \STATE Sample action $\mathbf{a} \sim \pi_{\phi}(\cdot|\mathbf{s})$, execute $\mathbf{a}$, observe next state $\mathbf{s}'$, and calculate reward $r$.
                \STATE Store transition $(\mathbf{s},\mathbf{a},r,\mathbf{s}')$ in $\mathcal{D}$.
                \IF{$|\mathcal{D}| \geq B$}
                    \STATE Sample a mini-batch of $B$ transitions from $\mathcal{D}$.
                    \STATE For each transition, sample target action $\tilde{\mathbf{a}} \sim \pi_{\phi}(\cdot|\mathbf{s}')$.
                    \STATE Update critic networks $Q_{\theta_1}$ and $Q_{\theta_2}$ by minimizing:
                    $\mathcal{L}_{Q_i} = \frac{1}{B}\sum \Big( Q_{\theta_i}(\mathbf{s},\mathbf{a}) - \big( r + \gamma \big( \min_{j=1,2} Q_{\theta_j}(\mathbf{s}',\tilde{\mathbf{a}}) - \alpha \log \pi_{\phi}(\tilde{\mathbf{a}}|\mathbf{s}') \big) \big) \Big)^2$.
                    \STATE Update actor network $\pi_{\phi}$ by minimizing:
                    $\mathcal{L}_{\pi} = \frac{1}{B}\sum \big( \alpha \log \pi_{\phi}(\mathbf{a}|\mathbf{s}) - \min_{i=1,2} Q_{\theta_i}(\mathbf{s},\mathbf{a}) \big)$.
                    \STATE Update temperature $\alpha$ by minimizing:
                    $\mathcal{L}_{\alpha} = \frac{1}{B}\sum \big( -\alpha ( \log \pi_{\phi}(\mathbf{a}|\mathbf{s}) + \mathcal{H}_0 ) \big)$.
                    \STATE Soft update target networks: $\bar{\theta}_i \leftarrow \tau \theta_i + (1-\tau) \bar{\theta}_i$.
                \ENDIF
                \STATE $\mathbf{s} \leftarrow \mathbf{s}'$.
            \ENDFOR
        \ENDFOR
    \end{algorithmic}
\end{algorithm}
\section{Simulations}
In this section, we evaluated the performance of IGS in two-user RSMA systems via MATLAB simulations. Without loss of generality, the noise power was normalized to $\sigma^2 = 1$, and $|h_2| = 1$.

\begin{figure}[t!]
    \centering
    \centerline{\includegraphics[width=0.9\linewidth]{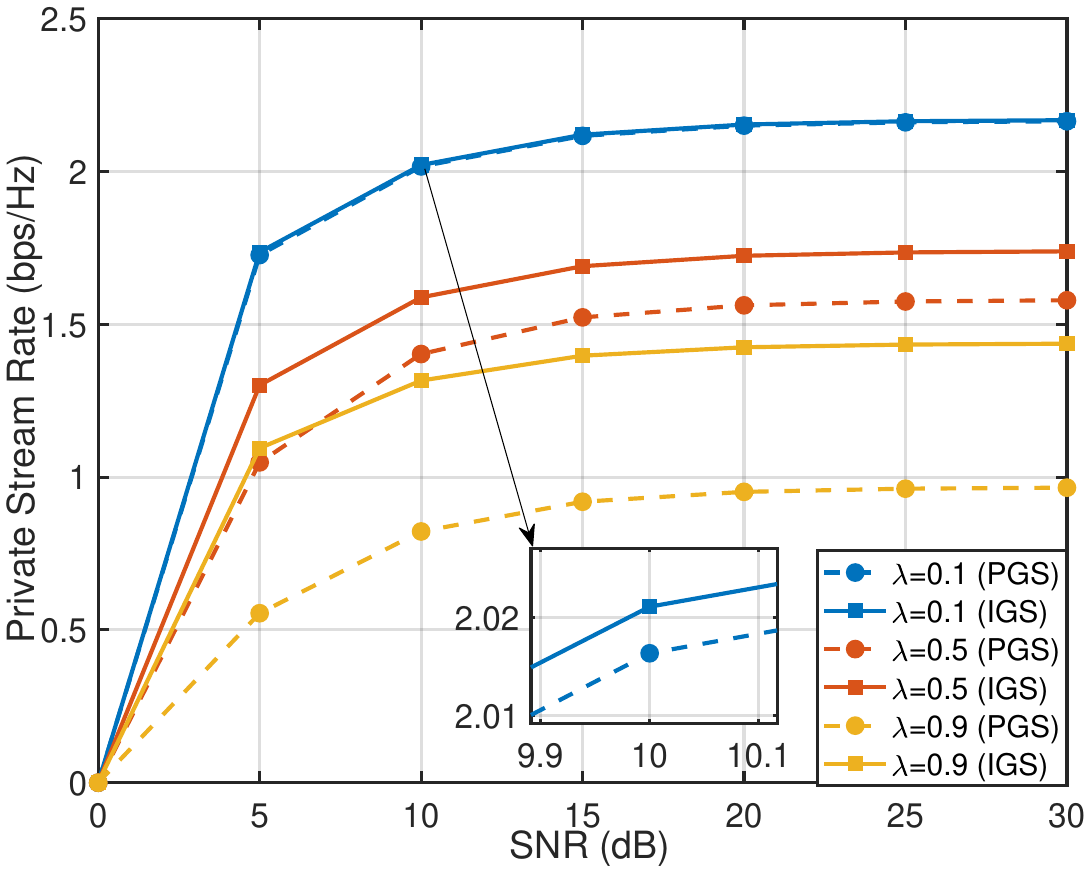}}
    \setlength{\abovecaptionskip}{0.cm}
    \caption{The sum of private rates as a function of the SNR.}
	\label{private_rate_vs_SNR}
\end{figure}

Fig.~\ref{private_rate_vs_SNR} plots the sum of private rates versus the SNR, with different values of the imperfect SIC coefficient $\lambda$ and the improper coefficient $\kappa$. The power threshold coefficient was fixed to be $\tau_{\text{SIC}} = 1$ and the channel gain ratio was set to $|h_1|/|h_2| = 5$. 
The results demonstrate that employing IGS enhanced the achievable private rate, and the performance gain over PGS was more significant as the level of SIC imperfection $\lambda$ increased.

\begin{figure}[t!]
    \vspace{-0.2cm} 
    \centering
    \centerline{\includegraphics[width=0.9\linewidth]{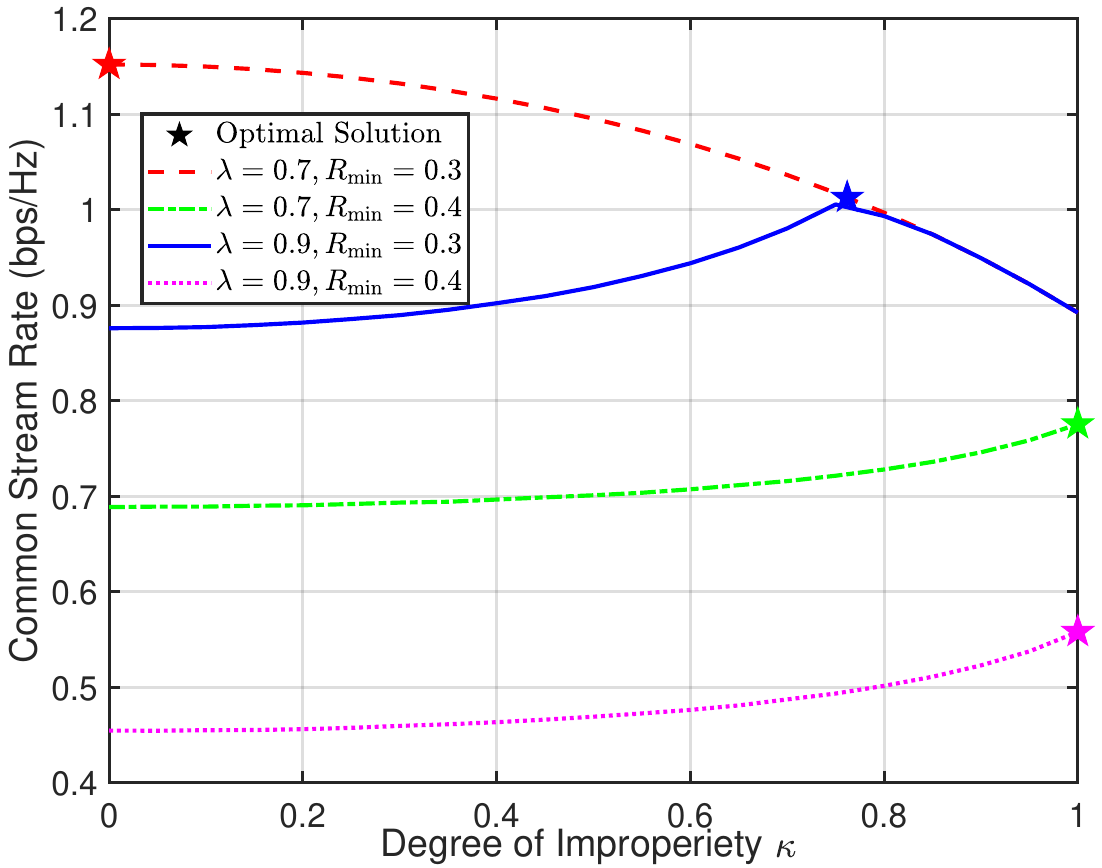}}
    \setlength{\abovecaptionskip}{0.cm}
    \caption{The common rate as a function of the degree of impropriety $\kappa$ with different $\lambda$ and $R_{\min}$.}
	\label{common_rate_vs_impropriety}
\end{figure}

We next evaluated the common rate $\ R_c$ as a function of the degree of impropriety $\kappa$ for different SIC imperfection coefficients $\lambda$ and minimum rate constraints $R_{\min}$, shown in Fig.~\ref{common_rate_vs_impropriety}. The system parameters were set as $p_1 = p_2 = 1.7$, $|h_1|/|h_2| = 2$, and $\tau_{\text{SIC}} = 2$. The theoretically optimal $\kappa$ were obtained from \eqref{kappa_optimal}, highlighted with stars in Fig.~\ref{common_rate_vs_impropriety}, while the common power $p_c$ for each $\kappa$ was optimized through the numerical search. The results show that the monotonicity of  $R_c$ with respect to $\kappa$ depended on both $\lambda$ and $R_{\min}$, in agreement with the theoretical analysis in Section \ref{Common Streams}. 
These findings demonstrate that employing IGS and jointly optimizing its degree of impropriety can effectively improve the common rate performance of RSMA systems.
%
%
\begin{table}[t!]
    \centering
    \caption{Parameters of the Soft Actor-Critic algorithm}
    \label{tab:sac_parameters}
    \resizebox{0.45\textwidth}{!}{
    \belowrulesep=0pt 
    \aboverulesep=0pt
    \renewcommand{\arraystretch}{1}
    \begin{tabular}{|l|c|l|c|}
    \toprule
    \textbf{Parameter} & \textbf{Value} & \textbf{Parameter} & \textbf{Value}  \\
    \midrule
    Learning Rate (Actor/Critic) & $3 \times 10^{-4}$ & Temperature ($\alpha$) & 0.2   \\
    \hline
    Target Update Rate ($\tau$) & 0.005 & Discount Factor ($\gamma$) & 0.99  \\
    \hline
    Replay Buffer Size & $1 \times 10^{6}$ &  Batch Size & 256  \\
    \hline
    Hidden Layer Size & [256, 256] & Target Entropy & $-\text{dim}(\mathcal{A})$ \\
    \bottomrule
    \end{tabular}
    }
\end{table}

\begin{figure}[t!]
    \centering
    \centerline{\includegraphics[width=0.9\linewidth]{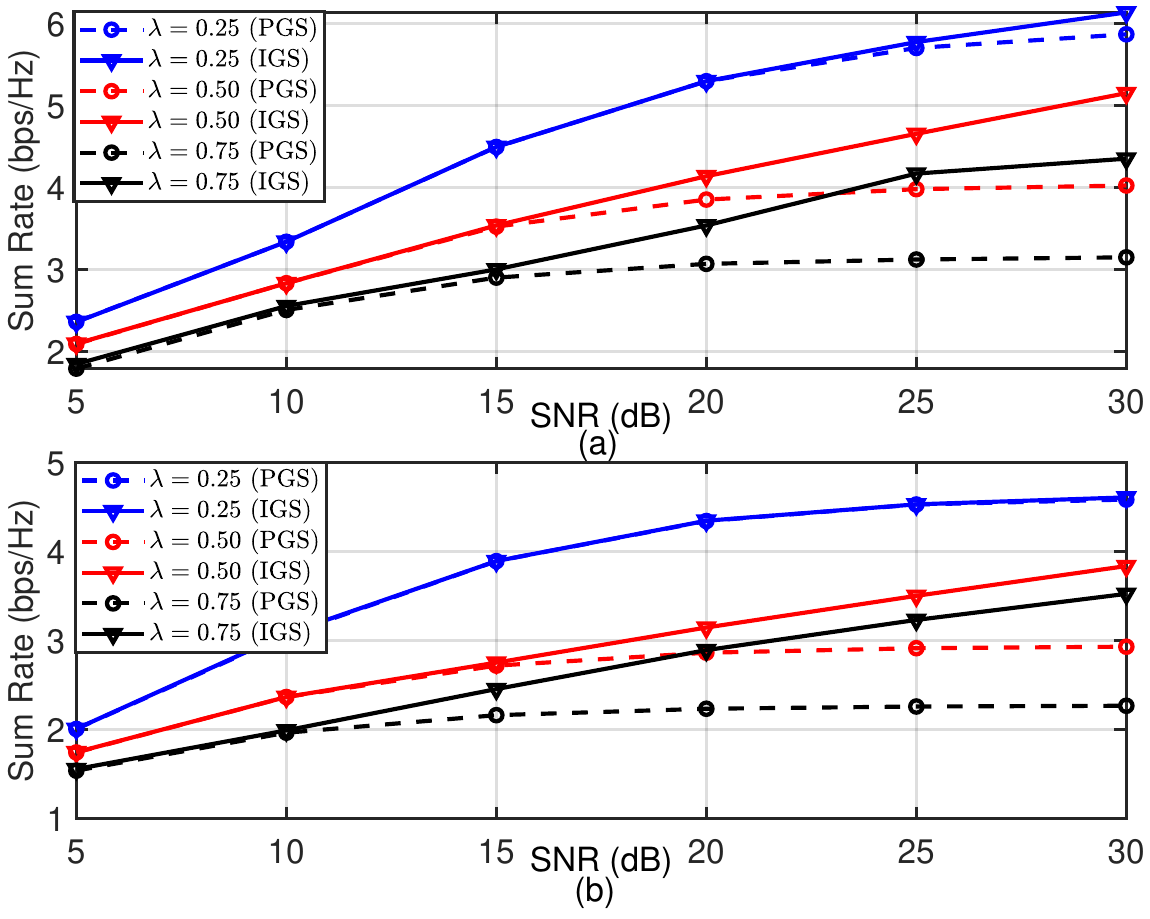}}
    \setlength{\abovecaptionskip}{0.cm}
    \caption{The optimized sum rate as a function of the SNR with different $\lambda$. The lower bound was (a) $R_{\min}$= 0.2, and (b) $R_{\min}$= 0.5.}
	\label{exp3_sumrate_diff_Rmin}
\end{figure}

\begin{figure}[!t]
    \centering
    \centerline{\includegraphics[width=0.9\linewidth]{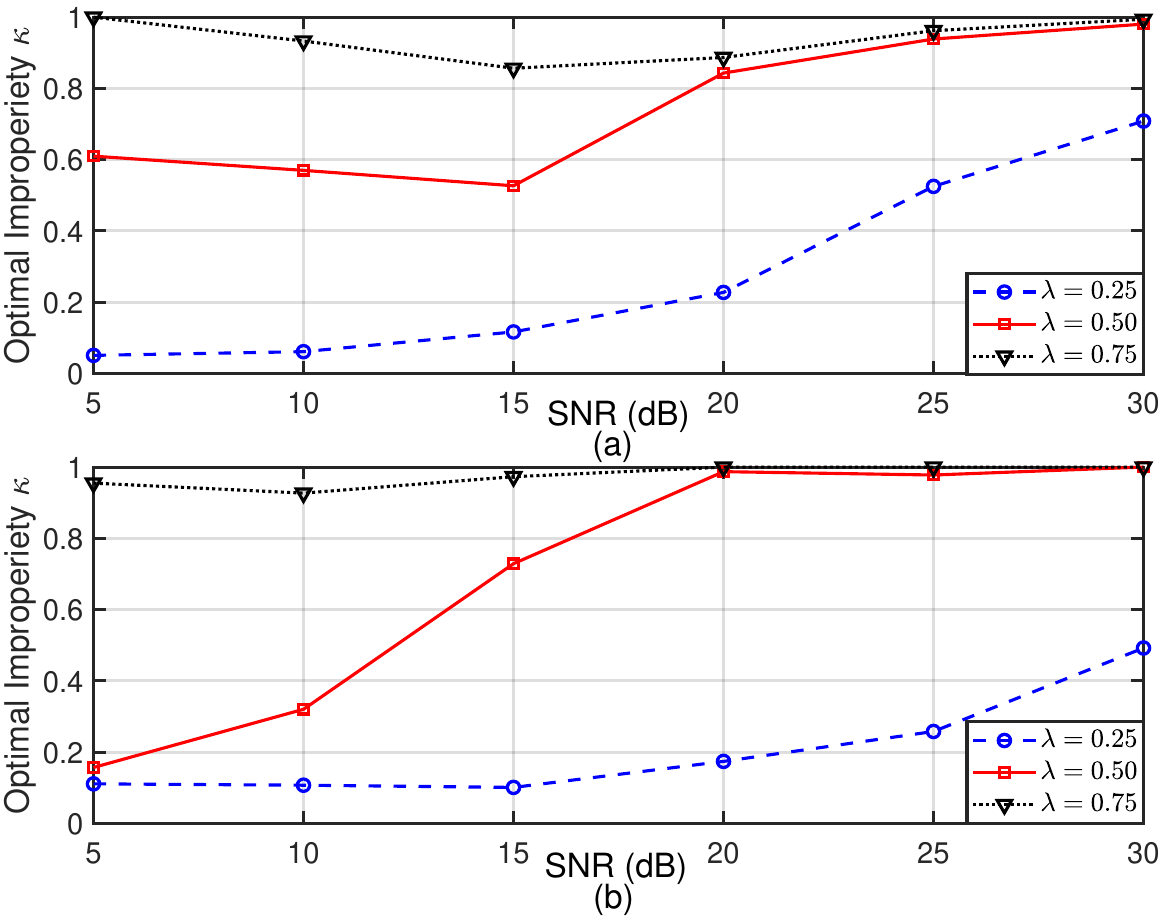}}
    \caption{The optimal $\kappa$ for the sum rate maximization as a function of the SNR. The lower bound was (a) $R_{\min}=0.2$, and (b) $R_{\min}=0.5$.}
	\label{exp3_sumrate_optimal_kappa_Rmin}
\end{figure}

The third experiment employed the SAC algorithm, whose parameters were specified in Table~\ref{tab:sac_parameters}, to solve the non-convex sum rate maximization problem in \eqref{eq:Max_sumrate}. 
Fig.~\ref{exp3_sumrate_diff_Rmin} compares the achievable sum rate of RSMA systems with either PGS or IGS, for different minimum rate requirements $R_{\min}$ and a power threshold $\tau_{\text{SIC}}=1$. The results show that IGS outperformed PGS in the high SNR regime, and this performance advantage became more significant for a lower $R_{\min}$. Fig.~\ref{exp3_sumrate_optimal_kappa_Rmin} further illustrates the optimal impropriety $\kappa$ across different parameter configurations. The results indicate that a higher SIC imperfection generally required a higher degree of impropriety $\kappa$, and the distinct behavior of $\kappa$ as a function of SNR for different constraints $R_{\min}$ can be attributed to the non-convex nature of the sum rate optimization problem in \eqref{eq:Max_sumrate} with strongly coupled system parameters.

%
%
\section{Conclusion}
We have investigated the optimization of the achievable rate in downlink RSMA systems enhanced by IGS in the presence of imperfect SIC. 
Our framework has provided an analytically tractable approach to leverage IGS as an intrinsic signaling degree of freedom. Closed-form expressions for the optimal impropriety degree have been derived for both private and common rate maximization, while a SAC-based algorithm has been developed to tackle the non-convex sum rate problem. IGS consistently outperforms the conventional PGS across all evaluated scenarios, with the performance gap becoming increasingly pronounced as the level of SIC imperfection rises. These results suggest that the signal impropriety can be integrated in the design of future RSMA transceivers, particularly for practical deployments where the perfect interference cancellation cannot be guaranteed.

%

\bibliographystyle{IEEEbib}
\bibliography{yourbib}

@article{mao2018rate,
    title={Rate-splitting multiple access for downlink communication systems: {B}ridging, generalizing, and outperforming {SDMA} and {NOMA}},
    author={Y. Mao and B. Clerckx and V. O. Li},
    journal={EURASIP J. Wireless Com. Netw.},
    volume={2018},
    number={1},
    pages={133},
    year={2018},
    publisher={Springer}
}

@ARTICLE{Mao_Yijie_2022,
    title={Rate-Splitting Multiple Access: {F}undamentals, Survey, and Future Research Trends}, 
    author={Y. Mao and O. Dizdar and B. Clerckx and R. Schober and P. Popovski and H. V. Poor},
    journal={IEEE Commun. Surv. Tutor.}, 
    year={2022},
    volume={24},
    number={4},
    pages={2073-2126},
    doi={10.1109/COMST.2022.3191937}
}

@ARTICLE{Xu_Yongjun2024,
    title={Resource Allocation for {RSMA}-based Symbiotic Radio Systems under Imperfect {SIC} and {CSI}}, 
    author={Y. Xu and M. Wang and H. Zhang and Q. Xue and J. Kang and Q. Chen and C. Yuen},
    journal={IEEE Trans. Veh. Technol.}, 
    year={2024},
    volume={},
    number={},
    pages={1-6},
    doi={10.1109/TVT.2024.3492203}
}

@article{WANG2025102761,
    title = {Covert performance analysis in {IRS}-assisted {RSMA} systems with imperfect {SIC}},
    author = {S. Wang and B. Lian and M. Liu and J. Zhang and G. Huang and J. Sun and P. Qinn},
    journal = {Physical Commun.},
    volume = {72},
    pages = {102761},
    year = {2025},
    issn = {1874-4907},
}

@book{Schreier_Book,
    title       = {Statistical Signal Processing of Complex-Valued Data: {T}he Theory of Improper and Noncircular Signals},
	author      = {P. J. Schreier and L. L. Scharf},
	publisher   = {Cambridge, U.K.: Cambridge Univ. Press},
	year        = {2010}
}

@ARTICLE{ZengYong_2013,
    title={Transmit Optimization With Improper {G}aussian Signaling for Interference Channels}, 
    author={Y. Zeng and C. M. Yetis and E. Gunawan and Y. L. Guan and R. Zhang},
    journal={IEEE Trans. Signal Process.}, 
    year={2013},
    volume={61},
    number={11},
    pages={2899-2913},
}

@ARTICLE{Cheng_Hao_2022,
    title={Improper {G}aussian Signaling for Downlink {NOMA} Systems With Imperfect Successive Interference Cancellation}, 
    author={H. Cheng and Y. Xia and Y. Huang and L. Yang},
    journal={IEEE Trans. Wireless Commun.}, 
    year={2022},
    volume={21},
    number={9},
    pages={7753-7763},
    doi={10.1109/TWC.2022.3161379}
}

@ARTICLE {Jin_Honglei_2025,
    title={Sum-Rate Maximization for Uplink Multi-User {NOMA} with Improper {G}aussian Signaling: {A} Deep Reinforcement Learning Approach}, 
    author={H. Jin and Z. Li and H. Cheng and Y. Xia and H. Hu},
    journal={IEEE Trans. Veh. Technol.}, 
    year={2025},
    volume={},
    number={},
    pages={1-13},
}

@ARTICLE{Yang_Zhaohui_2021,
    title={Optimization of Rate Allocation and Power Control for Rate Splitting Multiple Access ({RSMA})}, 
    author={Z. Yang and M. Chen and W. Saad and M. Shikh-Bahaei},
    journal={IEEE Trans. Commun.}, 
    year={2021},
    volume={69},
    number={9},
    pages={5988-6002},
    doi={10.1109/TCOMM.2021.3091133}
}

@ARTICLE{Soleymani_2023_TVT,
    title={Rate Splitting in {MIMO} {RIS}-Assisted Systems With Hardware Impairments and Improper Signaling}, 
    author={M. Soleymani and I. Santamaria and E. A. Jorswieck},
    journal={IEEE Trans. Veh. Technol.}, 
    year={2023},
    volume={72},
    number={4},
    pages={4580-4597},
}

@ARTICLE{Salem_2023_TWC,
    title={Secure Rate Splitting Multiple Access: {H}ow Much of the Split Signal to Reveal?},
    author={A. Salem and C. Masouros and B. Clerckx},
    journal={IEEE Trans. Wireless Commun.},
    year={2023},
    volume={22},
    number={6},
    pages={4173-4187},
    doi={10.1109/TWC.2022.3223961}
}

@InProceedings{haarnoja2018SAC,
    title = 	 {Soft {A}ctor-{C}ritic: {O}ff-policy Maximum Entropy Deep Reinforcement Learning with a Stochastic Actor},
    author =       {T. Haarnoja and A. Zhou and P. Abbeel and S. Levine},
    booktitle = 	 {Proc. 35th Int. Conf. Mach. Learn.},
    pages = 	 {1861-1870},
    year = 	 {2018},
    volume = 	 {80},
}

@ARTICLE{Xu_TCOM_2026,
  author={Xu, D. and Zhu, H.},
  journal={IEEE Trans. Commun.}, 
  title={Hybrid {RSMA} Systems With Improper {G}aussian Signaling Under Imperfect {SIC}}, 
  year={2026},
  volume={74},
  number={},
  pages={2268-2283},
  }

@ARTICLE{Soleyman_TGCN_2022,
  author={Soleymani, M. and Santamaria, I. and Schreier, P. J.},
  journal={IEEE Trans. Green Commun. Netw.}, 
  title={Improper Signaling for Multicell {MIMO} {RIS}-Assisted Broadcast Channels With {I}/{Q} Imbalance}, 
  year={2022},
  volume={6},
  number={2},
  pages={723-738},
  }
\end{document}